\documentclass[aps,preprint]{revtex4}
\usepackage{amsfonts}
\usepackage{amssymb}
\usepackage{amsmath}
\usepackage{graphicx}
\usepackage{hyperref}
\usepackage{float}
\usepackage{placeins}
\usepackage[font={footnotesize,it}]{caption}
\usepackage{latexsym}
\usepackage{xcolor}
\usepackage{hyperref}
\usepackage{orcidlink}
\usepackage{bm} 
\begin{document}
\title{Lorentz Symmetry Violation in Charged Black Hole Thermodynamics and Gravitational Lensing: Effects of the Kalb-Ramond Field}

\author{Mert Mangut \orcidlink{0000-0003-3364-1923}}
\email{mert.mangut@emu.edu.tr}
\affiliation{Physics Department, Eastern Mediterranean
University, Famagusta, 99628 North Cyprus via Mersin 10, Turkey.}

\author{Huriye G\"{u}rsel \orcidlink{0000-0002-6531-5156}}
\email{huriye.gursel@emu.edu.tr}
\affiliation{Physics Department, Eastern Mediterranean
University, Famagusta, 99628 North Cyprus via Mersin 10, Turkey.}

\author{\.{I}zzet Sakall{\i} \orcidlink{0000-0001-7827-9476}}
\email{izzet.sakalli@emu.edu.tr}
\affiliation{Physics Department, Eastern Mediterranean
University, Famagusta, 99628 North Cyprus via Mersin 10, Turkey.}

\begin{abstract}
This study investigates the consequences of Lorentz symmetry violation in the thermodynamics and gravitational lensing of charged black holes coupled to the Kalb-Ramond field. We first explore the impact of Lorentz-violating parameters on key thermodynamic properties, including the Hawking temperature, entropy, and specific heat, demonstrating significant deviations from their Lorentz-symmetric counterparts. Our analysis reveals that the Lorentz-violating parameter $b$ induces modifications in phase transitions and stability conditions, offering novel insights into black hole thermodynamics. Additionally, the influence of Lorentz symmetry breaking on gravitational lensing is examined using modifications to the Rindler-Ishak method, showing that these effects enhance the bending of light near compact objects. Our findings, derived within the framework of the standard model extension and bumblebee gravity models, suggest that Lorentz-violating corrections could lead to observable astrophysical phenomena, providing potential tests for deviations from Einstein's theory of relativity.
\end{abstract}

\maketitle

\section{Introduction} \label{isec1}
The concept of Lorentz symmetry, fundamental in modern physics, asserts that physical laws remain consistent across different inertial reference frames. While extensively supported by experimental evidence, various theoretical frameworks suggest that under specific energy conditions, Lorentz symmetry may deviate \cite{isKostelecky:2003fs,isColladay:1998fq}. These frameworks include string theory \cite{isMaldacena:1997re}, loop quantum gravity \cite{isAharony:1999ti,isBirrell:1982ix}, Horava--Lifshitz gravity \cite{isHorava:2009uw}, non-commutative field theory \cite{isConnes:1997cr,isSeiberg:1999vs}, Einstein-aether theory \cite{isJacobson:2007veq,isJacobson:2010mx}, massive gravity \cite{isDeser:1981wh}, $f(T)$ gravity \cite{isSaridakis:2017eqb}, $f(R,T)$ gravity \cite{isHarko:2011kv}, $f(R,T,L_M)$ \cite{isHaghani:2021fpx}, very special relativity \cite{isCohen:2006ky}, and others \cite{isPalti:2019pca,isHorndeski:2024sjk}.

The breakdown of Lorentz symmetry occurs in two ways: explicitly and spontaneously \cite{isKostelecky:1988zi}. Explicit breaking involves the absence of Lorentz invariance in the Lagrangian density, leading to different physical laws in certain frames. On the other hand, spontaneous breaking happens when the Lagrangian density maintains Lorentz invariance, but the system's ground state does not exhibit Lorentz symmetry \cite{isHiggs:1964pj,isNilles:1983ge}. Moreover, the exploration of spontaneous Lorentz symmetry breaking is rooted in the Standard Model Extension \cite{isColladay:1998fq}, where bumblebee models encapsulate the simplest field theories. In these models, a bumblebee field's non-zero vacuum expectation value violates local Lorentz invariance, impacting thermodynamic properties and other phenomena \cite{isBertolami:2005bh,isKanzi:2019gtu,isKanzi:2021cbg,isMangut:2023oxa,isUniyal:2022xnq,isKanzi:2022vhp}.

Recent studies have focused on bumblebee gravity, analyzing solutions for static and spherically symmetric spacetime akin to Schwarzschild and (Anti-)de Sitter-Schwarzschild configurations \cite{isref1711.02273,isref2011.12841}. The introduction of rotating bumblebee black holes and their properties, such as Hawking radiation, greybody factors, accretion processes, and quasinormal modes, has expanded this research domain \cite{isOliveira:2021abg,isMalik:2023bxc,isJha:2023vhn,refBumblebeeFoundational1, refBumblebeeFoundational2, refBumblebeeRecent1,refBumblebeeRecent2,refBumblebeeRecent3,refBumblebeeRecent4,refBumblebeeRecent5,refBumblebeeRecent6,refBumblebeeRecent7,refBumblebeeRecent8}. Additionally, the Kalb-Ramond (KR) field \cite{isref2308.06613}, coupled with gravity, has been explored for its spontaneous Lorentz symmetry breaking effects \cite{isGogoi:2022wyv}, leading to investigations into gravitational lensing, quasinormal modes, greybody factors, shadows cast, and the dynamics of particles near the KR black holes \cite{isGuo:2023nkd,refKRFoundational1, refKRRecent1, refKRRecent2,refKRRecent3,refKRRecent4,refKRRecent5}.

Gravitational wave studies, particularly in black hole physics, have gained prominence with advancements in detection technology like LIGO and VIRGO \cite{isLIGOScientific:2016aoc,isKAGRA:2013rdx,isMiddleton:2024ytu}. Recent studies have yielded novel exact solutions for charged static and spherically symmetric spacetime, both with and without a cosmological constant, within the background of the KR field's non-zero vacuum expectation value \cite{isDuan:2023gng}. In this research, we will refer to these black holes as CBHwKRF (charged black holes within the KR field), which have recently been the focus of significant research \cite{isZahid:2024ohn,isZahid:2024hyy,isLiu:2024axg,isLiu:2024lve,isHosseinifar:2024wwe,isGuo:2023nkd}.

Advances in black hole thermodynamics have explored the impact of quantum corrections on black hole entropy \cite{isFaulkner:2013ana,isHarlow:2014yka,isSen:2012dw,isMathur:2009hf}. It has been reported that the black hole entropy is modified by a corrected term due to non-perturbative quantum effects. These corrections significantly affect the black hole mass and other thermodynamic quantities, especially for small black holes. For instance, the Schwarzschild black hole mass decreases due to quantum corrections, and the stability of 4D Schwarzschild and Schwarzschild-AdS black holes is influenced at small areas \cite{isPourhassan:2024yfg,isPourhassan:2020yei}. The thermodynamics and statistics of black holes are analyzed by computing the partition function and deriving conditions to satisfy the Smarr–Gibbs–Duhem \cite{isDehghani:2008qr} relation in the presence of these quantum corrections \cite{isPradhan:2016feg}.

In this paper, we investigate the effects of Lorentz symmetry breaking by studying the thermodynamics and gravitational lensing of the CBHwKRF. It is worth highlighting that the static, spherically symmetric charged black hole solution in the presence of the KR field, as examined in this work, was originally derived in Ref. \cite{isDuan:2023gng}. In this manuscript, we aim to explore the novel consequences of Lorentz symmetry violation on quantum-corrected thermodynamic properties and gravitational lensing, building on this established solution. The paper is organized as follows: In Sec. \ref{isec2}, we present the CBHwKRF spacetime and derive the relevant field equations. We also discuss the first law of thermodynamics and Smarr's formula for this spacetime. Section \ref{isec3} analyzes thermal fluctuations with quantum corrections in the CBHwKRF spacetime. In Sec. \ref{isec4}, we explore gravitational lensing with a background of the Lorentz-breaking effect. Section \ref{isec5} presents relevant applications in astrophysics. Finally, Sec. \ref{isec6} summarizes our results and discussion.

{\color{black}
\section{CBH\lowercase{w}KRF spacetime} \label{isec2}
In this study, we focus on exploring a static and spherically symmetric spacetime with a non-zero vacuum expectation value (VEV) for the KR field \cite{isDuan:2023gng}. The action of the gravity theory under consideration is explicitly given as \cite{Altschul:2009ae,Lessa:2019bgi}

\begin{multline}
S= \frac{1}{2} \int d^4 x \sqrt{-g}\left[R-2 \Lambda-\frac{1}{6} H^{\mu v \rho} H_{\mu v \rho}-V\left(B^{\mu v} B_{\mu \nu} \pm b^2\right)\right. \\
\left.+\xi_2 B^{\rho \mu} B^v{ }_\mu R_{\rho v}+\xi_3 B^{\mu v} B_{\mu \nu} R\right]+\int d^4 x \sqrt{-g} \mathcal{L}_{\mathrm{M}},
\end{multline}

where $H_{\mu v \rho}$ is the KR field strength, $\Lambda$ represents the cosmological constant, and $\xi_{2,3}$ denotes the non-minimal coupling constants between gravity and the KR field ($B_{\mu \nu}$), which is a rank-two antisymmetric tensor field satisfying $B_{\mu \nu} = -B_{\nu \mu}$. It is worth noting that $8 \pi G=1$ for simplicity. The matter Lagrangian $\mathcal{L}_{\mathrm{M}}$ corresponds to the electromagnetic field, expressed as $\mathcal{L}_{\mathrm{M}}=-\frac{1}{2} F^{\mu \nu} F_{\mu \nu}+\mathcal{L}_{\text {int }}$, with $F_{\mu \nu}=\partial_\mu A_\nu-\partial_\nu A_\mu$ (the reader is referred to the specifics outlined in Ref. \cite{isDuan:2023gng}).

The CBHwKRF spacetime metric is described by \cite{isDuan:2023gng}
\begin{equation} \label{metric}
 \mathrm{d}s^2 = -A(r) \mathrm{d}t^2 + B(r) \mathrm{d}r^2 + r^2 \mathrm{d}\theta^2 + r^2 \sin^2 \theta \mathrm{d}\phi^2,   
\end{equation}

where $A(r)$ and $B(r)$ are radial-dependent metric functions reflecting the KR field's impact with the pseudo-electric field $\tilde{E}(r)$
\begin{equation}
\tilde{E}(r) = |\ell| \sqrt{\frac{A(r) B(r)}{2}},    
\end{equation}
in which $\ell$ is associated with the constant norm condition $b_{\mu\nu} b^{\mu\nu} = -b^2$ as $b=\xi_2\ell^2 / 2$. Here, $\ell_{\mu\nu}$ represents the background antisymmetric tensor field associated with the KR field, which is responsible for spontaneous Lorentz symmetry breaking \cite{isDuan:2023gng}.  After making straightforward calculations, $A(r)={B(r)}^{-1}$ has been found to be 
\begin{equation} \label{ismf}
A(r) = \left( \frac{1}{1-b} - \frac{2M}{r} + \frac{q^{2}}{(1-b)^{2}r^{2}}\right),
\end{equation}
where $q$ is the electric charge. The Lorentz-violating parameter $b$ is constrained by gravitational experiments, and as $b \to 0$, we recover the Reissner-Nordstr\"{o}m case \cite{isStephani:2003tm}. The horizons are given by
\begin{equation} \label{horizon}
r_{\pm} = (1-b)\left( M \pm \sqrt{M^{2} - \frac{q^{2}}{(1-b)^{3}}} \right).
\end{equation}
When \(q \to 0\), these results align with previous studies \cite{isCasana:2017jkc,isOvgun:2018ran,isSakalli:2023pgn}.  Fig. \ref{fig01} and Fig. \ref{fig02} illustrate the behaviors of the horizons under different parameters and values. The plots of both figure demonstrate how the Lorentz-violating parameter \( b \) and the charge \( q \) affect the horizon formation of the CBHwKRF. Higher values of \( b \) reduce the horizon radius showing that Lorentz violation plays crucial role in determining the horizons.

\begin{figure}[H]
 \hspace*{-13mm}\begin{tabular}{@{}cccc@{}}
    \includegraphics[scale=0.9]{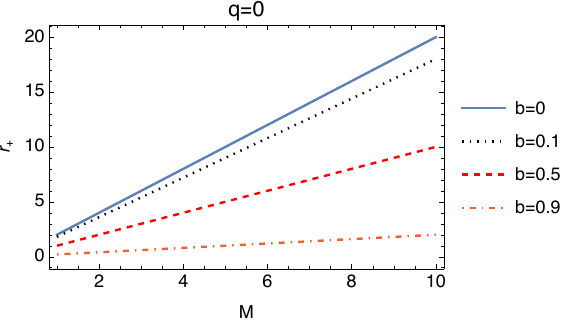} &
   \includegraphics[scale=0.9]{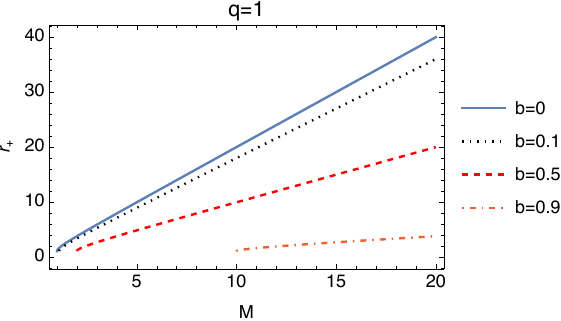}
  \end{tabular}
  \caption{Event horizon ($r_+$) versus mass graph. The plots are governed by Eq. \eqref{horizon}.} \label{fig01}
\end{figure}

\begin{figure}[H]
 \hspace*{-13mm}\begin{tabular}{@{}cccc@{}}
    \includegraphics[scale=0.9]{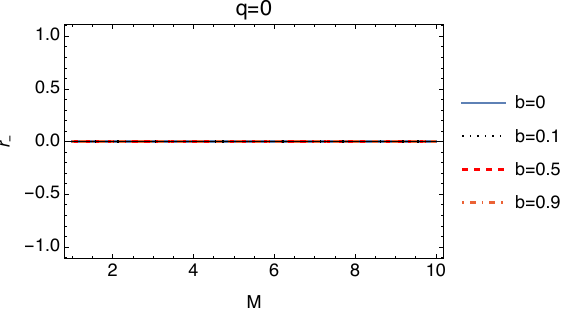} &
    \includegraphics[scale=0.9]{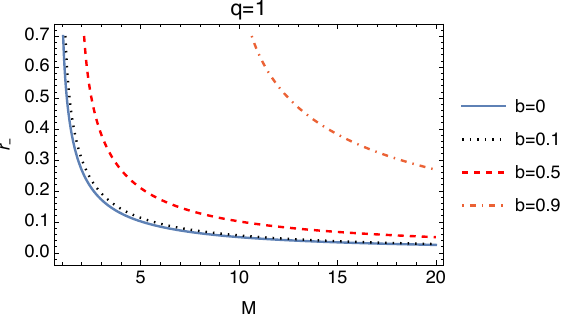}
    \end{tabular}
  \caption{Inner or Cauchy horizon ($r_-$) versus mass graph. The plots are governed by Eq. \eqref{horizon}.} \label{fig02}
\end{figure}

-eps-converted-to.pdf
The following exact differential form can define the thermodynamic structure of the thermal system, which is the so-called first law of thermodynamics of black holes \cite{isWei:2009zzc,istHooft:1986vqu}:

\begin{equation} \label{firstlaw}
dM = T_{H}dS + \Phi dq,
\end{equation}

where $T_{H}$ is the black hole or Hawking temperature, $S$ is the black hole or the so-called Hawking-Bekenstein entropy, and $\Phi$ is nothing but the electric potential given in Eq. \ref{pot}. If we put $r=r_{+}$ into the metric function, the metric function goes to zero and the  mass of  the  black  hole can be calculated as 

\begin{equation}
M(q,r_{+})=\frac{r_+}{2(1-b)}+\frac{q^{2}}{2(1-b)^2r_+}.\label{m5}
\end{equation}

In geometric units ($G=c=\hbar=1$), the Hawking-Bekenstein entropy for standard spherical symmetric black holes is given by \cite{isBekenstein:1973ur}

\begin{equation}
S=\pi r^{2}_{+}.\label{m6}
\end{equation}

Therefore, one can rewrite Eq. \eqref{m5} as follows

\begin{equation}
M(S,q)=\frac{S^{1/2}}{2\pi^{1/2}(1-b)}+\frac{q^{2}\pi^{1/2}}{2(1-b)^2S^{1/2}}.\label{m7}
\end{equation}

Now we can use the homogeneous function theorem of Euler for finding the Smarr's formula \cite{isKastor:2009wy}. According to Euler's homogeneous theorem, the two variables homogeneous function of order $n$ is given by 

\begin{equation}
f(\lambda^{i}x,\lambda^{j}y)=\lambda^{n}f(x,y) \label{mm7}
\end{equation}

where $\lambda$ is a  constant  and $(i,j,k)$ are integer powers \cite{m}. For the case, the differentiation of Eq. \eqref{mm7}  becomes

\begin{equation}
f(x,y)=n^{-1} \bigg[i\frac{\partial f}{\partial x}x+j\frac{\partial f}{\partial y}y\bigg].
\end{equation}

When we choose $(i=2n, j=n)$ and define the new shifted variables as $(S \rightarrow \lambda^{i} S, q \rightarrow \lambda^{j} q)$, the Smarr's  formula is constructed as follows using Eq. \eqref{m7} and Eq. \eqref{mm7}

\begin{equation}
M(S,q)=2S\left(\frac{\partial M}{\partial S}\right)+q\left(\frac{\partial M}{\partial q}\right).\label{m9}
\end{equation}

To find the Hawking temperature, one can isolate $T_{H}$ from Eq. \eqref{firstlaw}:

\begin{equation}
T_{H} = \left( \frac{\partial M}{\partial S} \right),
\end{equation}

which results in

\begin{equation}
T_{H}=\frac{1}{4S^{1/2}\pi^{1/2}(1-b)}-\frac{q^{2}\pi^{1/2}}{4(1-b)^2S^{3/2}},\label{m10}
\end{equation}

or

\begin{equation}
T_{H}=\frac{(1-b)r_+^2-q^2}{4\pi(1-b)^2r_+^3}.\label{mm10}
\end{equation}

On the other hand, one can check the result by considering the surface gravity $\kappa$, which is calculated from the derivative of the metric function at the event horizon \cite{isWald:1984rg}:
\begin{equation}
f'(r) = \frac{d}{dr} \left( \frac{1}{1-b} - \frac{2M}{r} + \frac{q^2}{(1-b)^2 r^2} \right),
\end{equation}
where prime symbol denotes the derivative with respect to $r$. Evaluating this at $r = r_{+}$, we have
\begin{equation}
f'(r_+) = \frac{2M}{r_+^2} - \frac{2 q^2}{(1-b)^2 r_+^3}.
\end{equation}
Substituting the expression for $M$:
\begin{equation}
M = \frac{1}{2} \frac{r_+}{1-b} + \frac{1}{2} \frac{q^2}{(1-b)^2 r_+},
\end{equation}
we get
\begin{equation}
f'(r_+) = \frac{1}{(1-b) r_+} +\frac{q^2}{(1-b)^2 r_+^3}.
\end{equation}

The surface gravity ($\xi$) of a spherically symmetric static metric is given by  \cite{isWald:1984rg}
\begin{equation}
\xi = \frac{1}{2} f'(r_+).
\end{equation}

Thus,
\begin{equation}
\xi = \frac{1}{2} \left( \frac{1}{(1-b) r_+} - \frac{q^2}{(1-b)^2 r_+^3} \right).
\end{equation}

Since the Hawking temperature ($T_H$) \cite{isWald:1984rg} is defined by
\begin{equation}
T_H = \frac{\kappa}{2\pi},
\end{equation}
one can obtain
\begin{equation}
T_H = \frac{1}{4\pi} \left( \frac{1}{(1-b) r_+} - \frac{q^2}{(1-b)^2 r_+^3} \right),
\end{equation}
which is fully agree with the result obtained in Eq. \ref{mm10}.

The subsequent derivative in Smarr's formula specifies the electric potential energy at the horizon. For this reason, the electric potential interaction is represented by
\begin{equation}
\Phi= \left(\frac{\partial M}{\partial q}\right)=\frac{q\pi^{1/2}}{(1-b)^2S^{1/2}}=\frac{q}{(1-b)^2r_+}. \label{pot}
\end{equation}
In addition, the electric field of CBH\lowercase{w}KRF spacetime is given by

\begin{equation}
  \mathcal{E}= - \nabla \Phi   \hat{r}=\frac{q}{2(1-b)^2r_+^2} \hat{r}.
\end{equation}

To sum up, the compact Smarr's expression can be written as

\begin{equation} \label{smarr}
M=2ST_{H}+q\Phi.
\end{equation}

Also, we can calculate the heat capacity \cite{isDebnath:2019yit,isSoroushfar:2023acx}, which plays a key role in the analysis of thermal stability, using the formula below

\begin{equation} C_H=T_{H}\left(\frac{\partial S}{\partial T_{H}}\right), \label{s25}
\end{equation}

thus, one can find the heat capacity as follows

\begin{equation}
C_H=\frac{2 \pi  r^2 \big((1-b) r^2-q^2\big)}{(b-1) r^2+3 q^2}. \label{s26}
\end{equation}

\begin{figure}[H] 
\centering
 \hspace*{-13mm} \begin{tabular}{@{}cccc@{}}
    \includegraphics{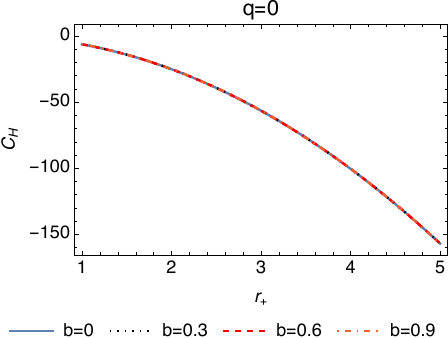} &
                                \\
    \includegraphics{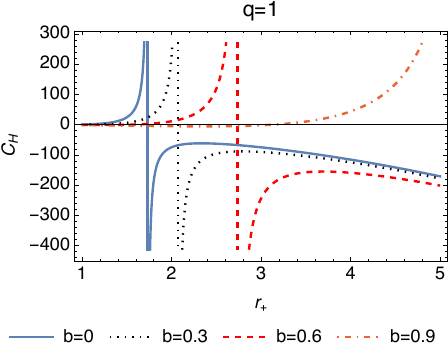} &
                                \\
    \end{tabular}
  \caption{The plots of $C_{H}$ are governed by Eq. \eqref{s25}.} \label{fig1}
\end{figure}

The plots presented in Fig. \ref{fig1} illustrate the behavior of the quantity $C_H$ as a function of $r$, under different settings of the parameters $q$ and $b$. Each subplot corresponds to a unique value of $q$, showing the variation in $C_H$ with $r_+$ for different values of $b$. For $q = 0$, the system remains thermally stable with no divergences. As $q$ increases, divergences emerge (becoming more pronounced at higher $q$) while increasing $b$ mitigates these effects by smoothing the transitions and shifting critical points to larger $r_+$. This suggests that $b$ plays a stabilizing role in the thermodynamic behavior of the system.

Our analysis focuses on the outer spacetime of the black hole, relevant for astrophysical observations. For $q > q_{\text{crit}} = \sqrt{(1-b)^3}M$, a naked singularity forms, but we restrict our study to regimes where the outer horizon exists, enabling the examination of thermodynamic and lensing properties. The charged black hole's thermodynamic behavior with Lorentz-symmetry violation shows critical phenomena in $C_H$, signaling second-order phase transitions. The Helmholtz free energy $F$ \cite{isSadeghi:2016dvc}, however, does not exhibit similar transitions, likely due to the influence of $b$. This suggests a modified thermodynamic framework where conventional phase-transition behavior may not apply. The existence of critical points in $C_H$ underscores the system's complexity, warranting future investigation for deeper insights.
}

{\color{black}
\section{Thermal Fluctuations in CBH\lowercase{w}KRF Spacetime} \label{isec3}

Standard thermodynamic analysis, supplemented by corrections from statistical mechanics, allows us to examine thermal fluctuations in the structures considered \cite{nnn17, nnnn17}. These corrections lead to notable variations in the classical thermodynamic potentials. In this context, if the system is in thermal equilibrium, the density of states is given by \cite{nn17}
\begin{equation}
\rho(E)=\frac{e^{S_{0}}}{\sqrt{2\pi}}\left[\left(\frac{\partial^2S(\beta)}{\partial\beta^2}\right)_{\beta=\beta_0} \right]^{1/2},    
\end{equation}
where $S_0$ represents the uncorrected entropy, and $\beta_0=1/T_H$. The formula for the logarithmic corrected (LC) entropy is given as
\begin{equation}
S^{LC}=S_{0}-\frac{\alpha}{2} \ln (S_0T^{2}_{H}) + (\text{sub-leading terms}). \label{s39}
\end{equation}
Here, $S_{0}=\pi r^{2}_{+}$ and $\alpha$ is the parameter representing thermal fluctuations \cite{nn17}. Notably, $\alpha=1$ corresponds to the maximum effect of thermal fluctuations \cite{nn17,n17}. Substituting Eq. \eqref{mm10} and Eq. \eqref{s26} into Eq.\eqref{s39}, the LC entropy of the CBHwKRF spacetime can be expressed as
\begin{equation} 
S^{LC}=\pi r_+^2-\frac{\alpha}{2} \ln\left(\frac{\left((1-b) r_+^2-q^2\right)^2}{16 \pi  (1-b)^4 r_+^4}\right). \label{mm11}
\end{equation} 

\begin{figure}[H] 
\centering
\hspace*{-13mm} \begin{tabular}{@{}cccc@{}}
    \includegraphics{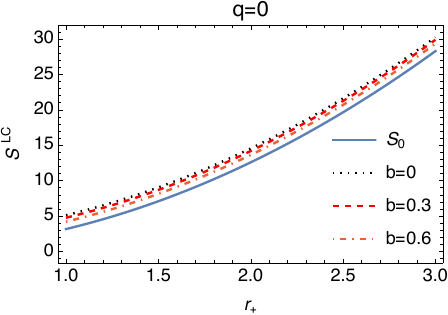} &
                                \\
    \includegraphics{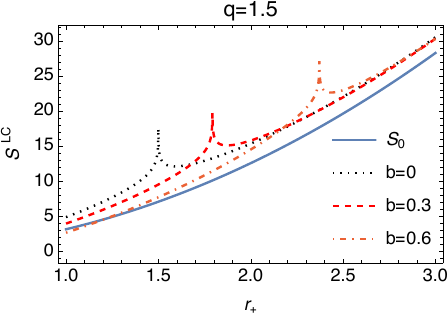} &
                                \\
\end{tabular}
\caption{The plots of $S^{LC}$ are governed by Eq. \eqref{mm11} under the influence of maximum thermal fluctuations $(\alpha=1)$. } \label{fig2}
\end{figure}

The plots provided by Fig. \ref{fig2} illustrate the behavior of the quantity $S^{LC}$ as a function of the event horizon $r_+$, for various settings of the parameters $q$ and $b$. Each plot is indicative of how $S^{LC}$ responds to changes in $r_+$ under different parameter configurations. As $q$ increases, the behavior of $S^{LC}$ transitions from linear and predictable (for $q = 0$) to complex and non-linear (for $q = 1.5$ ). Higher $q$ values introduce intricate interactions and abrupt changes, highlighting the sensitivity of the system’s entropy to both $r_+$ and $b$. This suggests a critical role for $q$ in driving the thermodynamic complexity of the black hole. With this new expression for entropy, we can analyze significant thermodynamic energies and expressions under the influence of maximum thermal fluctuations $(\alpha=1)$. First, let us calculate the internal energy formulated as
\begin{equation}
E^{LC}=\int T_{H} dS^{LC}. \label{s41}
\end{equation}
Upon substituting $(T_H)$ and the differential of $(S^{LC})$ into Eq. \eqref{s41} and integrating, the internal energy is determined as
\begin{equation} 
E^{LC}=\frac{q^2 \left(3 \pi  r_+^2+1\right)+3 \pi  (1-b) r_+^4}{6 \pi  (1-b)^2 r_+^3}. \label{ss2}
\end{equation}

\begin{figure}[!ht] 
\centering
\hspace*{-13mm} \begin{tabular}{@{}cccc@{}}
    \includegraphics{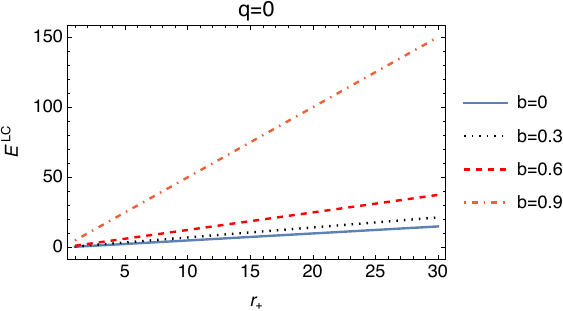} &
                                \\
    \includegraphics{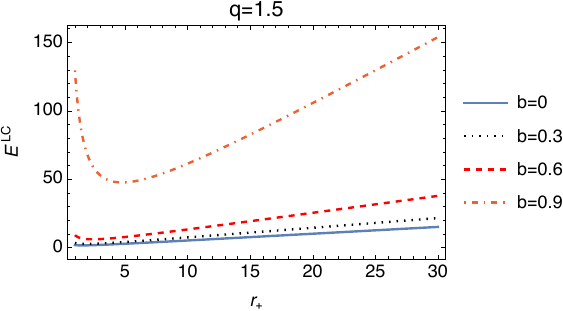} &
                                \\
\end{tabular}
\caption{The plots of $E^{LC}$ are governed by Eq. \eqref{ss2}. } \label{fig3}
\end{figure}

The graphs of Fig. \ref{fig3} illustrate the variation of a physical quantity, $E^{LC}$, with respect to the event horizon radius $r_+$, for different values of charge parameter $q$ and parameter $b$. As $q$ increases, the behavior of $E^{LC}$ shifts from linear and stable (for $q = 0$) to non-linear and complex (for $q = 1.5$ ). Higher $q$ values highlight intricate interactions and a stronger influence of $b$, especially as $E^{LC}$ transitions to non-monotonic or decreasing trends with $r_+$.


The expression for the corrected Helmholtz free energy $(F^{LC})$ is given by
\begin{equation}
F=-\int S^{LC}dT_{H}. \label{s43}
\end{equation}
Integrating Eqs. \eqref{mm10} and \eqref{mm11} into Eq. \eqref{s43}, the corrected Helmholtz free energy can be formulated as
\begin{equation}
\begin{aligned}
&F^{LC}=\frac{1}{24 \pi (1-b)^2 r_+^3} \bigg[ 
3 \left((1-b) r_+^2 - q^2 \right) \ln \left( \frac{\left((1-b) r^2 -q^2 \right)^2}{(1-b)^4 r_+^4} \right) \\
&\quad + 3 (1-b) r_+2 \left(2 \pi r_+^2 -\ln (16 \pi )  \right) + q^2 \left( 18 \pi r_+^2 + 4 + 3 \ln (16 \pi ) \right) 
\bigg]. \label{mmm2}
\end{aligned}
\end{equation}

\begin{figure}[H]
\centering
\hspace*{-13mm} \begin{tabular}{@{}cccc@{}}
    \includegraphics{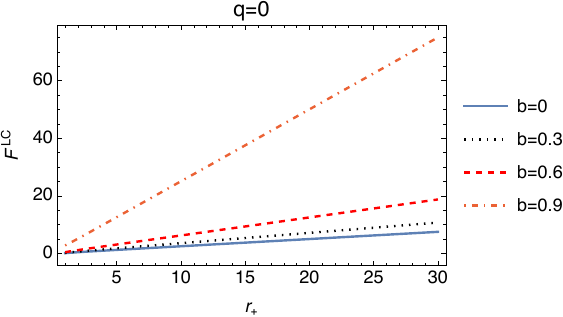} &
                                \\
    \includegraphics{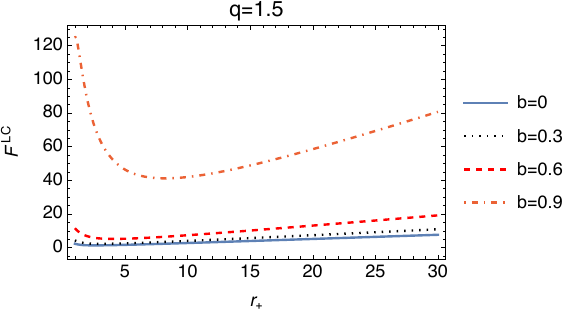} &
                                \\
\end{tabular}
\caption{The plots of $F^{LC}$ are governed by Eq. \eqref{mmm2}.} \label{fig4}
\end{figure}

The plots of Fig. \ref{fig4} represent the variation of the quantity $F^{LC}$ with respect to the radius of the event horizon $r$, for different values of the parameters $q$ and $b$. For $q = 0$, $F^{LC}$ exhibits a linear increase with $r_+$, maintaining a stable dependence on $b$. As $q$ increases, particularly at $q = 1.5$, this growth slows, with higher $b$ values causing a plateau or slight decline. Overall, increasing $b$ mitigates these fluctuations, underscoring its crucial role in governing the thermodynamic behavior of the system.

Referring to the definition of Helmholtz free energy $(F=E-TS)$, the pressure of the black hole is expressed as
\begin{equation}
P=\frac{dF}{dV}, \label{mmm1}
\end{equation}
where $V=\frac{4}{3}\pi r^{3}_{H}$, represents the volume of the black hole. Substituting Eq. \eqref{mmm2} into Eq. \eqref{mmm1} and deriving the expression for pressure, we obtain
\begin{equation} \label{plc}
P^{LC}=\frac{\left((1-b) r_+^2-3 q^2\right) \left(-\ln\left(\frac{\left((1-b) r_+^2-q^2\right)^2}{(1-b)^4 r_+^4}\right)+2 \pi  r_+^2+\ln (16 \pi )\right)}{32 \pi ^2 (1-b)^2 r_+^6}.
\end{equation}

\begin{figure}[H]
\centering
\hspace*{-13mm} \begin{tabular}{@{}cccc@{}}
    \includegraphics{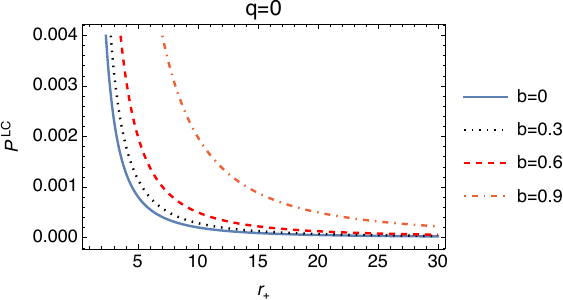} &
                                \\
    \includegraphics{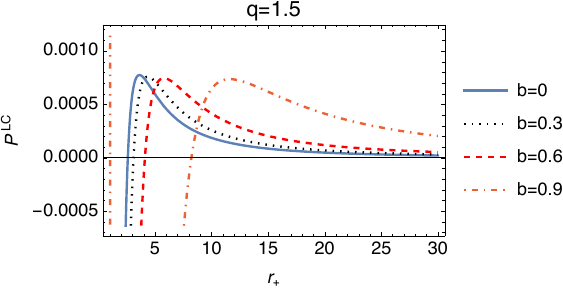} &
                                \\
\end{tabular}
\caption{The plots of $P^{LC}$ are governed by Eq. \eqref{plc}. } \label{fig5}
\end{figure}

The plots given in Fig. \ref{fig5} illustrate the behavior of the physical quantity \( P_{LC} \) as a function of event horizon radius \( r_+\), parameterized by charge \( q \) and another variable \( b \). For $q=0$, $P_{LC}$ decreases monotonically with $r_+$, stabilizing as it approaches zero. At $q$ increases, a peak emerges for certain $b$ values, followed by a decline, indicating more complex interactions. For example at $q=1.5$, the behavior becomes increasingly varied, with $P_{LC}$ crossing zero or exhibiting chaotic fluctuations, highlighting significant dependencies on both $b$ and $r_+$. Overall, these trends reveal intricate dynamics influenced by the parameters $q$ and $b$.

Furthermore, the enthalpy $(H)$ is defined as follows:
\begin{equation}
H=E+PV. \label{s35}
\end{equation}
By substituting the corresponding values of internal energy, pressure, and volume into Eq. \eqref{s35}, we derive
\begin{equation} \label{hlc}
\begin{aligned}
H^{LC}=&-\frac{\left((1-b) r_+^2 - 3 q^2\right) \ln\left(\frac{\left((1-b) r_+^2 - q^2\right)^2}{(1-b)^4 r_+^4}\right)}{24 \pi (1-b)^2 r_+^3} + \frac{(1-b) r_+^2 \left(14 \pi r_+^2  + \ln (16\pi)\right)}{24 \pi (1-b)^2 r_+^3} \\
&+ \frac{q^2 \left(6 \pi r_+^2 + 4 - 3 \ln (16 \pi)\right)}{24 \pi (1-b)^2 r_+^3}.
\end{aligned}
\end{equation}

\begin{figure}[H] 
\centering
\hspace*{-13mm} \begin{tabular}{@{}cccc@{}}
    \includegraphics{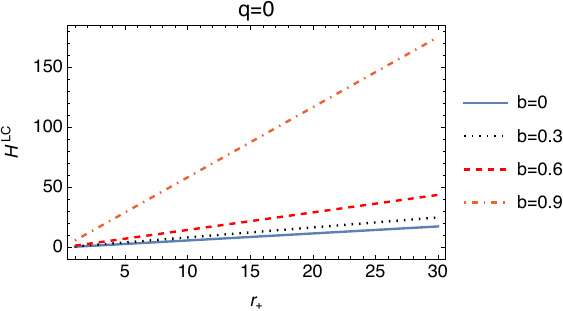} &
                                \\
    \includegraphics{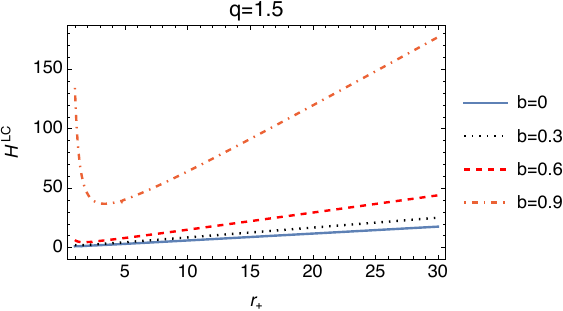} &
                                \\
\end{tabular}
\caption{The plots of $H^{LC}$ are governed by Eq. \eqref{hlc}. } \label{fig6}
\end{figure}

As illustrated in Fig. \ref{fig6}, for $q=0$, $H_{LC}$ displays a linear relationship with $r_+$, consistent across all $b$ values. With increasing $q$, such as $q=1.5$, linearity continues but with minor slope differences influenced by $b$, where higher $b$ values result in a more gradual rise. These variations do not disrupt the primary proportionality between $H_{LC}$ and $r_+$, as $b$ mainly modulates the growth rate without altering the fundamental trend.

The thermodynamic expression for Gibbs free energy $(G)$ can be written as
\begin{equation}
G=F+PV. \label{s37}
\end{equation}
When the values of $F^{LC}$, $P^{LC}$, and $V$ are substituted into Eq. \eqref{s37}, the Gibbs free energy reduces to
\begin{equation} \label{glc}
G^{LC}=\frac{(1-b) r_+^2 \ln \left(\frac{\left((1-b) r_+^2-q^2\right)^2}{(1-b)^4 r_+^4}\right)+(1-b) r^2 \left(4 \pi  r_+^2-\ln (16 \pi )\right)+q^2 \left(6 \pi  r_+^2+2\right)}{12 \pi  (1-b)^2 r_+^3}.
\end{equation}

\begin{figure}[H] 
\centering
\hspace*{-13mm} \begin{tabular}{@{}cccc@{}}
    \includegraphics{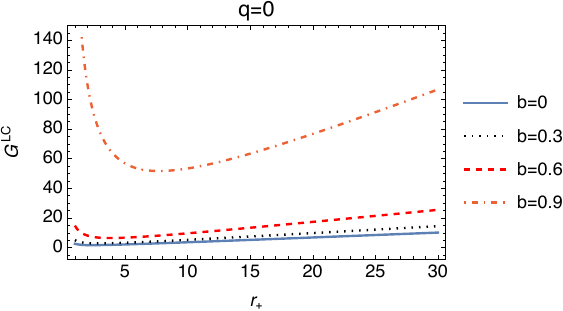} &
                                \\
    \includegraphics{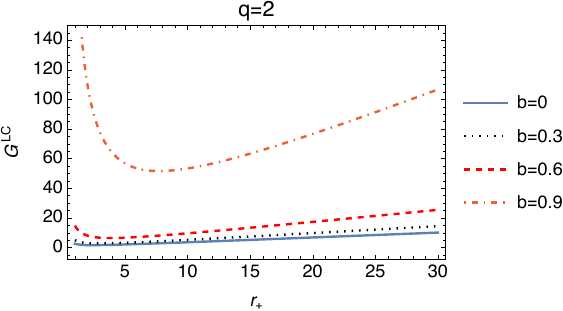} &
                                \\
\end{tabular}
\caption{The plots of $G^{LC}$ are governed by Eq. \eqref{glc}. } \label{fig7}
\end{figure}

The plots of $G_{LC}$ in Fig. \ref{fig7} show a consistent increase with $r_+$ across all $q$ values, with $q = 0$ starting at a lower value and exhibiting a steeper slope compared to higher $q$ values, such as $q=2$. As $b$ increases, the initial value of $G_{LC}$ at $r_+ = 5$ decreases, while the growth rate remains largely unaffected, indicating that $b$ primarily shifts the baseline without altering the trend. Overall, $G_{LC}$ demonstrates a strong dependence on $r_+$, with $b$ modulating its initial value across different charge scenarios.

The corrected specific heat $(C^{LC})$ can be described as
\begin{equation}
C^{LC}=\frac{dE^{LC}}{dT}.
\end{equation}

Using Eq. \eqref{ss2} and Eq. \eqref{mm10}, we can compute the expression governing the corrected specific heat of black holes. Consequently, the corrected specific heat is expressed as
\begin{equation} \label{clc}
C^{LC}=\frac{2 \left(\pi  (1-b) r_+^4-q^2 \left(\pi  r_+^2+1\right)\right)}{3 q^2-(1-b) r_+^2}.
\end{equation}

\begin{figure}[H]
\centering
\hspace*{-13mm} \begin{tabular}{@{}cccc@{}}
    \includegraphics{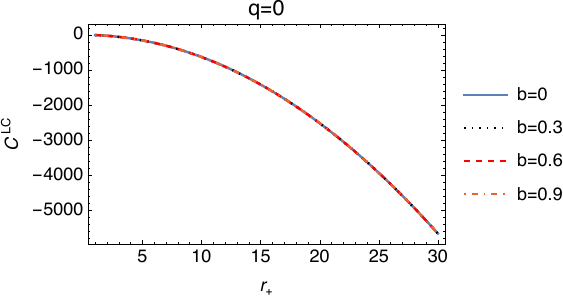} &
                                \\
    \includegraphics{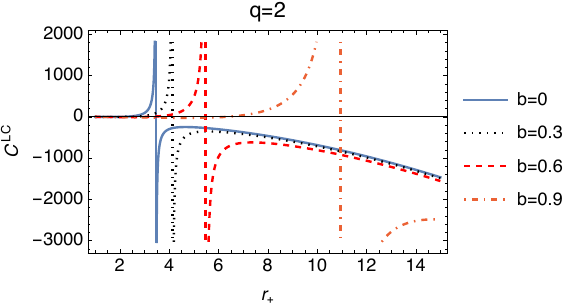} &
                                \\
\end{tabular}
\caption{The plots of $C^{LC}$ are governed by Eq. \eqref{clc}. }  \label{fig8}
\end{figure}
}

The plots in Fig. \ref{fig8} show that for $q=0$, $C^{LC}$ decreases smoothly with increasing $r_+$, indicating predictable thermal behavior. At higher $q$ values (i.e., $q=2$), the behavior becomes more complex, with fluctuations and divergences near specific $r_+$ values. Increasing $b$ stabilizes these fluctuations, moderating the abrupt changes and shifting critical points to higher $r_+$, highlighting its role in improving the thermal stability of the system. Moreover, Fig. \ref{fig8} highlights the sensitivity of $C_{LC}$ to the thermal fluctuation parameter $\alpha$ alongside $q$ and $b$. For $q=2$, higher $b$ values shift the divergence point of $C_{LC}$ to larger $r_+$, indicating that the instability occurs in larger black holes. This emphasizes the role of $b$ in modulating thermal properties and stability.

The isothermal compressibility, which plays an important role in stability, is defined as

\begin{equation}
\kappa=-\frac{1}{V}\frac{\partial V}{\partial p} \bigg\vert_T. \label{k}
\end{equation}

Additionally, it should be noted that the condition $\kappa\geq 0$ ensures that the system returns to equilibrium under spontaneous changes of the parameters, and this is known as Le Chatelier's principle \cite{w}. When we put Eq. \eqref{plc} into Eq. \eqref{k} for using chain rule, the isothermal compressibility can be written as

\begin{equation}
\begin{aligned}
 \kappa = & \; 48 \pi ^2 (1-b)^2 r_+^6 \left(q^2-(1-b) r_+^2\right)\left\{q^2 r_+^2(1-b)  \left(14 \pi  r_+^2-2+11 \ln (16 \pi )\right) \right.\\
 &\left. +\left((1-b) r_+^2-q^2\right) \left(2 (1-b) r_+^2-9 q^2\right) \ln \left(\frac{\left((1-b) r_+^2-q^2\right)^2}{(1-b)^4 r_+^4}\right) \right.\\
 &\left. - 2 (1-b)^2 r_+^4 \left(\pi  r_+^2  + \ln (16\pi )\right) - 3 q^4 \left(4 \pi  r_+^2 - 2 + 3 \ln (16 \pi )\right) \right\}^{-1}. \label{k1}
\end{aligned}
\end{equation}

\begin{figure}[H]
\centering
\hspace*{-13mm} \begin{tabular}{@{}cccc@{}}
    \includegraphics{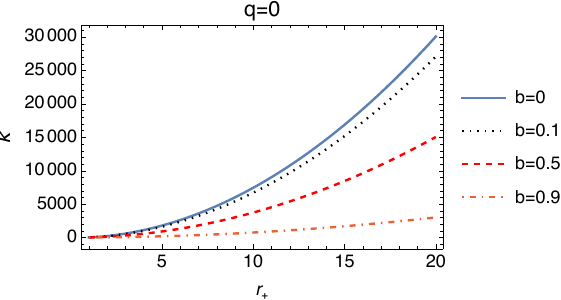} &
                                \\
    \includegraphics{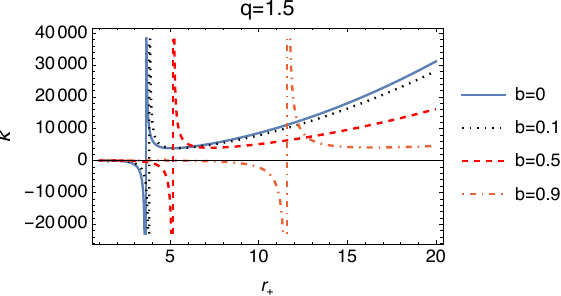} &
                                \\
\end{tabular}
\caption{Plots of the isothermal compressibility with respect to the event horizon. The graphs are governed by Eq. \eqref{k}. } \label{fig111}  
\end{figure}

Figure \ref{fig111} illustrates the behavior of the isothermal compressibility $\kappa$ as a function of $r_+$. For $q = 0$, $\kappa$ remains positive and increases with $r_+$ across all $b$ values, indicating stable thermodynamic behavior. In contrast, for higher $q$ values such as $q = 1.5$, $\kappa$ exhibits regions of negativity, particularly for higher $b$ values, signaling instability in certain parameter ranges. Additionally, divergence points appear, highlighting regions of extreme compressibility. This suggests that both $q$ and $b$ play crucial roles in determining stability, with higher values leading to greater instability.

In summary, our thermodynamic analysis notably advances LSV studies, particularly concerning charged black holes in KR field models. Unlike previous works that predominantly explore uncharged black holes or simplified LSV contexts, our study investigates the interplay between the LSV parameter $b$ and the charge $Q$, revealing unique fluctuation behaviors in thermodynamic quantities such as the Hawking temperature $T$, heat capacity $C_P$, and Helmholtz free energy $F$. The critical transitions observed in $C_P$ demonstrate a distinct dependence on $b$, setting our findings apart from other modified gravity scenarios. Additionally, our work integrates gravitational lensing effects using the Gauss-Bonnet theorem to establish an observational link to LSV. The explicit deflection angle calculations provide insights into the astrophysical manifestations of the KR field, offering a complementary perspective to the thermodynamic results. This dual approach—bridging black hole thermodynamics and astrophysical observables—enhances the significance of our analysis. Thus, our study contributes to the literature by extending LSV effects to charged black holes in KR field models, uncovering novel thermodynamic characteristics, and providing an observational context through lensing analysis.

{\color{black}

\section{Gravitational Lensing with a Background of the Lorentz-Breaking Effect} \label{isec4}

One of the renowned predictions of Einstein's theory of relativity is the gravitational lensing effect, which provides experimental verification of the theory. Rindler and Ishak (RI) investigated a lensing analysis method using the generalization of the inner product in a Riemannian manifold \cite{1}. In the symmetry plane ($\theta=\pi/2$), the generic RI lensing geometry structure can be illustrated as follows:

\begin{figure}[H]
\centering
\includegraphics[width=120mm,scale=1]{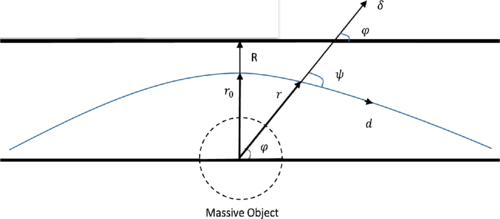}
\caption{The general bending of light geometry of the RI method}
\end{figure}

The lensing invariant between the path of light ($d$) and the radial line ($\delta$) as shown in Fig. 1 is given by

\begin{equation}
\cos(\psi) = \frac{d^{i}\delta_{i}}{\sqrt{(d^{i}d_{i})(\delta^{j}\delta_{j})}} = \frac{g_{ij}d^{i}\delta^{j}}{\sqrt{(g_{ij}d^{i}d^{j})(g_{kl}\delta^{k}\delta^{l})}}, \label{1}
\end{equation}

where $g_{ij}$ is the metric tensor of the spacetime. The geometrical coordinates of the lensing geometry are defined by

\begin{eqnarray}
d &=& (dr,d\varphi) = (A,1) d\varphi \text{, \ \ \ \ \ } d\varphi < 0, \notag \\
\delta &=& (\delta r,0) = (1,0) \delta r, \label{2}
\end{eqnarray}

where $A(r,\varphi) \equiv \frac{dr}{d\varphi}.$ When we substitute the information from Eq. \eqref{2} into Eq. \eqref{1}, the invariant formula simplifies to

\begin{equation}
\tan(\psi) = \frac{\left[g^{rr}\right]^{1/2}r}{\left| A(r,\varphi) \right|}. \label{3}
\end{equation}

Consequently, the one-sided bending angle can be measured as $\epsilon = \psi - \varphi$. Moreover, we can calculate the light trajectory by solving the null geodesics equation. In this context, the RI lensing geometry line-element (optical metric) for standard spherically symmetric spacetimes at a constant time slice can be written as
\begin{equation}
dl^{2} = \frac{dr^{2}}{f(r)} + r^{2}d\varphi^{2}. \label{4}
\end{equation}

where $f(r)$ is the metric function of the given spacetime. If $E$ and $L$ represent energy and angular momentum, respectively, the constraints of the null geodesics equations are given by

\begin{equation}
\frac{dt}{d\tau} = -\frac{E}{f(r)}, \text{ \ \ \ \ \ } \frac{d\varphi}{d\tau} = \frac{L}{r^{2}}. \label{5}
\end{equation}

Here, $\tau$ is the proper time. Combining the constraints, the general null geodesics equation becomes

\begin{equation}
\left(\frac{dr}{d\varphi}\right)^{2} = \frac{r^{4}}{L^{2}}\left(E^{2} - \frac{L^{2}}{r^{2}}f(r)\right). \label{6}
\end{equation}

Defining the new variable $u = \frac{1}{r}$ ($u \ll 1$), the second-order differential form of Eq. \eqref{6} is

\begin{equation}
\frac{d^{2}u}{d\varphi^{2}} = -uf(u) - \frac{u^{2}}{2}\frac{df(u)}{du}. \label{7}
\end{equation}

Substituting the metric function into Eq. \eqref{7}, the null geodesics equation for the spacetime reads

\begin{equation}
\frac{d^{2}u}{d\varphi^{2}} + \beta u \approx 3 M u^2 - 2 q^2 u^3 \beta^2 + \mathcal{O}(u^4), \label{8}
\end{equation}

where $\beta = 1/(1-b)$. Equation \eqref{8} represents a second-order nonlinear differential equation; however, we can solve the equation using a well-known perturbative solution method. In this method, we consider the solution of the linear and homogeneous differential equation, $\frac{d^{2}u}{d\varphi^{2}} + \beta u = 0$, in the form of $u(\varphi) = \frac{\sin(\sqrt{\beta} \varphi)}{R}$. Note that $R$ is called the impact parameter and $R \gg 1$. When we substitute the solution of the linear and homogeneous differential equation into the nonlinear part of Eq. \eqref{8}, the perturbative solution of Eq. \eqref{8} is given by

\begin{equation}
\begin{aligned}
u(\varphi) =& \frac{\sin(\sqrt{\beta} \varphi)}{R} + \frac{M(\cos(2 \sqrt{\beta} \varphi) + 3)}{2\beta R^2} + \frac{3\varphi q^2\beta^{3/2}\cos(\sqrt{\beta} \varphi)}{4R^3} \\
& - \frac{\beta q^2(6 \sin(\sqrt{\beta} \varphi) + \sin(3 \sqrt{\beta} \varphi))}{16R^3} + O\left(\frac{1}{R}\right)^4. \label{9}
\end{aligned}
\end{equation}

Also, the derivative of $u(\varphi)$ with respect to $\varphi$ is

\begin{equation}
\begin{aligned}
A(r,\varphi) =& \left\{\frac{\beta (8 R^2 + 3 \beta q^2) \cos(\sqrt{\beta} \varphi) - 8 R M \sin(2 \sqrt{\beta} \varphi)}{8 R^3 \sqrt{\beta}} \right. \\
& \left. - \frac{3 \beta^{3/2} q^2 (4 \sqrt{\beta} \varphi \sin(\sqrt{\beta} \varphi) + \cos(3 \sqrt{\beta} \varphi))}{16 R^3} \right\}r^2. \label{10}
\end{aligned}
\end{equation}

If we choose $\theta=\pi/2$ in Eq.\eqref{9}, the parameter representing the minimum distance between a light ray and a massive object during gravitational lensing $r_0$, known as the closest approach distance, is found as

\begin{equation}
\begin{aligned}
\frac{1}{r_0} =& \frac{\sin(\sqrt{\beta} \pi/2)}{R} + \frac{M(\cos( \sqrt{\beta} \pi) + 3)}{2\beta R^2} + \frac{3\pi q^2\beta^{3/2}\cos(\sqrt{\beta} \pi/2)}{8R^3} \\
& - \frac{\beta q^2(6 \sin(\sqrt{\beta} \pi/2) + \sin(3 \sqrt{\beta} \pi/2))}{16R^3}. \label{90}
\end{aligned}
\end{equation}

When we set $\varphi = 0$ and consider $R \gg 1$, the dominant terms of Eq. \eqref{9} and Eq. \eqref{10} are found as
\begin{equation}
\begin{aligned}
&r = \frac{1}{u(\varphi = 0)} = \frac{R^2 \beta}{2M}, \\
&A(r,\varphi = 0) \approx \frac{r^2 \sqrt{\beta}}{R}. \label{11}
\end{aligned}
\end{equation}

In light of the information in Eq. \eqref{11}, Eq. \eqref{3} can be rewritten as 

\begin{equation}
\begin{aligned}
\tan(\epsilon) \approx \epsilon \approx \frac{2 M}{R \beta^{3/2}}\left\{1 + \frac{4 M^2 q^2}{\beta^3 R^4} - \frac{4 M^2}{R^2} \right\}^{1/2} \approx & \frac{2 M}{R \beta^{3/2}}\left\{1 - \frac{2 M^2}{R^2} + \frac{2 M^2 q^2}{\beta^3 R^4} \right\} \\
& + O\left(\frac{4M^5q^4}{\beta^{15/2}R^9}\right). \label{12}
\end{aligned}
\end{equation}
}

The parameter $b$, and whence $\beta$, which characterizes the extent of Lorentz symmetry violation, can indeed have observable consequences in the context of gravitational lensing. Although current astrophysical observables are not finely tuned to detect such small deviations from general relativity, future high-precision measurements, especially in the vicinity of compact objects such as supermassive black holes or neutron stars, could provide a way to constrain $b$. For example, gravitational lensing experiments, using techniques like very long baseline interferometry (VLBI) \cite{Ding:2024ktk}
 in the Event Horizon Telescope (EHT) \cite{EventHorizonTelescope:2019dse}, could detect minute deviations in the bending angle of light, which are sensitive to the Lorentz-violating effects encapsulated by $b$.

The influence of $b$ on the lensing angle (Eq. \ref{12}) suggests that larger deviations from zero would manifest as an enhancement of the lensing effect, particularly noticeable in regions of strong gravitational fields. This offers a potential pathway for using gravitational lensing as a probe for Lorentz symmetry violation in future astrophysical observations.

{\color{black}
\section{Possible Applications in Astrophysics about CBH\lowercase{w}KRF Spacetime and Constraints on Lorentz-Violating Parameter}} \label{isec5}
{\color{black}
Astrophysics continually benefits from the theoretical frameworks provided by general relativity, particularly in understanding phenomena such as gravitational lensing and redshift. This section explores significant astrophysical applications that rely on the principles of general relativity to explain the behavior of compact stars. We present empirical data and theoretical analyses that not only validate these principles, but also help in estimating key celestial parameters. Additionally, the impact of Lorentz violation parameters on these phenomena is examined, offering insights into their potential implications on the conventional understanding of spacetime dynamics.

\begin{table}[H]
\caption{Tabulated numerical values of mass, radius, and electric charge for compact stars expressed in solar masses ($M_{\odot}$) \cite{1a}.}
\label{table:nonlin}
\centering
\begin{tabular}{|l|c|c|r|}
\hline
Charged Compact Stars & Mass $(M_{\odot})$ & Radius (km) & Electric Charge (C) \\ \hline\hline
\;\;\;\;Vela X-1  & $1.77M_{\odot}$ & $9.56$ & $1.81\times 10^{20}$ \\ \hline
\multicolumn{1}{|c|}{SAXJ 1808.4-3658} & $1.435M_{\odot}$ & $7.07$ & $1.87\times 10^{20}$ \\ \hline
\;\;\;\;4U 1820-30  & $2.25M_{\odot}$ & $10$ & $1.89\times 10^{20}$ \\ \hline
\end{tabular}
\end{table}

In the graphical analysis, we used standard international units (S.I units) with conversion factors $Gc^{-2}$ and $G^{1/2}c^{-2}(4\pi \varepsilon _{0})^{-1/2}$ for mass and charge, respectively. Note that, $G=6.67408\times 10^{-11}m^{3}kg^{-1}s^{-2}$, $c=3\times 10^{8}ms^{-1}$, and $\varepsilon _{0}=8.85418\times 10^{-12}C^{2}N^{-1}m^{2}$.

In Fig. 12, the one-sided bending angles of the compact stars are plotted for the RN limit $(b=0)$ and the maximum value of the horizon condition $\bigg(1-b=\left(\frac{M}{q}\right)^{2/3}\bigg)$. For all compact stars, the Lorentz violation parameter $b$ plays a crucial role as the lensing effect increases dramatically when $b$ approaches $1-\left(\frac{M}{q}\right)^{2/3}$.

\begin{figure}[!ht] 
\centering
\begin{tabular}{@{}cccc@{}}
    \includegraphics{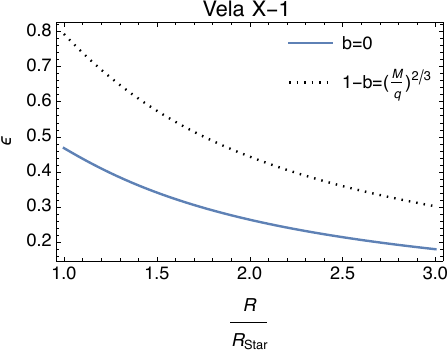} &
    \includegraphics{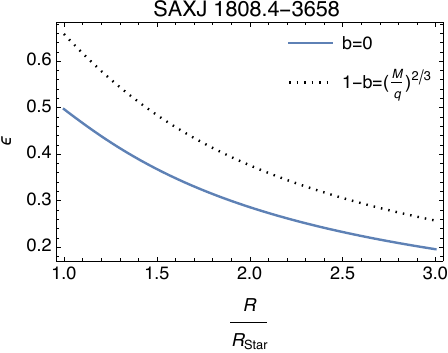} &
   \\
     \multicolumn{2}{c}{\includegraphics{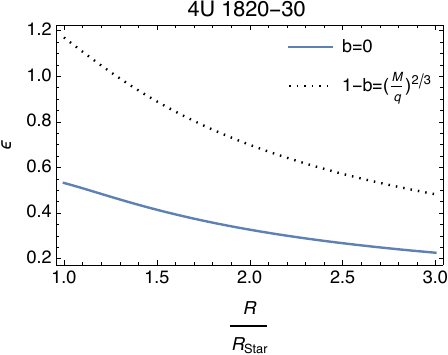}} 
  \end{tabular}
  \caption{These figures depict how the bending angle $\epsilon$ changes with respect to $R/R_{\text{star}}$ for the charged compact stars. Each graph compares the RN case ($b=0$) with the Lorentz violation effect ($b\neq0$). } \label{fig11}
\end{figure}

\newpage

Another intriguing astronomical effect from Einstein's theory of general relativity is the gravitational redshift. For static spacetimes, the gravitational redshift is given by 

\begin{equation}
z=\frac{\Delta\lambda}{\lambda} =\frac{\lambda_O}{\lambda_e}-1=\frac{1}{\sqrt{f(r)}}-1, \label{13}
\end{equation}

where $\lambda_O$ and $\lambda_e$ are the observed and emitted wavelengths, respectively \cite{1b,1c}. Substituting $f(r)$ into Eq. \eqref{13} under the assumption of large $r$ ($r \rightarrow R$ and $R>>1$), Eq. \eqref{13} simplifies to

\begin{equation}
z\approx -1+(1-b)^{1/2}+\frac{M(1-b)^{3/2}}{R}+\frac{q^2}{2(1-b)^{1/2}R^2}.
\label{14}
\end{equation}

In Fig. 13, we observe a similar graphical structure for gravitational redshift values of compact objects as in the bending of light analysis. In these graphs, the Lorentz violation contributes positively.

\begin{figure}[H]
\centering
\begin{tabular}{@{}cccc@{}}
    \includegraphics{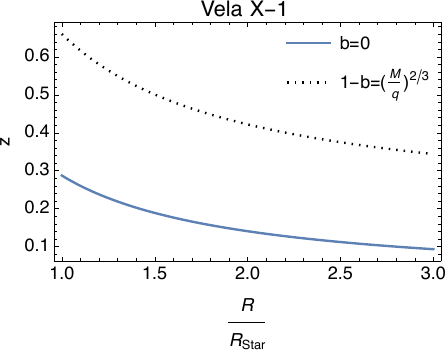} &
    \includegraphics{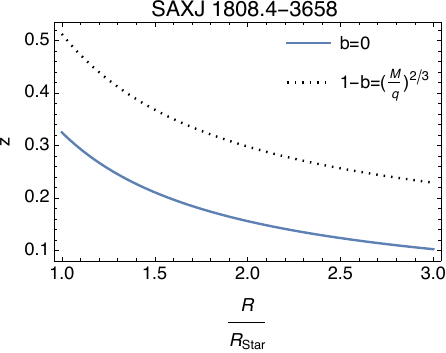} &
   \\
     \multicolumn{2}{c}{\includegraphics{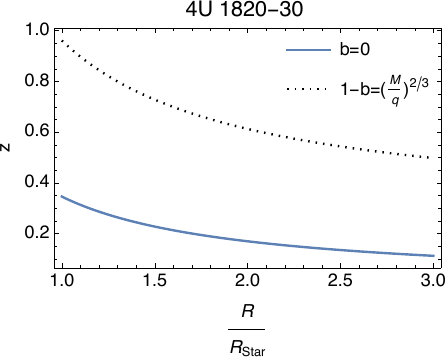}}
  \end{tabular}
  \caption{The graphs show how the gravitational redshift parameter $z$ evolves with respect to $R/R_{\text{star}}$ for the compact objects. Each graph includes trends for both the RN limit ($b=0$) and the Lorentz violation effect ($b\neq0$). }
\end{figure}
}

Finally, we aim to analyze potential constraints on $b$ related to thermal attributes and lensing observations. Given $b$'s influence on black hole thermodynamics and lensing, constraints are derived from criticality conditions in these areas. Typically, the thermodynamic stability criteria are used to derive the thermal threshold parameter $b_{\text{crit,thermal}}$ on $b$. For a CBHwKRF, this condition is linked to the behavior of the heat capacity $C_H$, which undergoes a divergence at a phase transition. The critical value of $b$ is determined by solving the condition for the divergence:

\begin{equation}
C_H=T \cdot \frac{\frac{\partial S}{\partial r_{+}}}{\frac{\partial T}{\partial r_{+}}}.
\end{equation}

The divergence occurs when $\frac{\partial T}{\partial r_{+}}=0$. Differentiating $T_H$ with respect to $r_+$, one can find
\begin{equation} \label{izs1}
\frac{\partial T_H}{\partial r_+} = \frac{- (1-b)r_+^4 + 3q^2r_+^2}{4\pi (1-b)^2 r_+^6}.
\end{equation}

Setting Eq. \eqref{izs1} to zero, we get the following condition:
\begin{equation}
- (1-b)r_+^4 + 3q^2r_+^2 = 0.
\end{equation}

Solving for $b_{\text{crit,thermal}}$ gives
\begin{equation}
b_{\text{crit,thermal}} = 1 - \frac{3q^2}{r_+^2}.
\end{equation}

The photon sphere condition \cite{Claudel:2000yi,Perlick:2004tq,Bozza:2002zj} is provided as
\begin{equation} \label{isps}
f(r_{\text{ph}}, b) - \frac{r_{\text{ph}} f'(r_{\text{ph}}, b)}{2} = 0,
\end{equation}
in which a prime symbol denotes the derivative with respect to $r$. Substituting Eq. \eqref{ismf}, $f(r, b))\equiv A(r) $, into Eq. \eqref{isps}, we have
\begin{equation}
\frac{1}{1-b} - \frac{2M}{r_{\text{ph}}} + \frac{q^2}{(1-b)^2r_{\text{ph}}^2} - \frac{r_{\text{ph}}}{2} \left( -\frac{2M}{r_{\text{ph}}^2} - \frac{2q^2}{(1-b)^2r_{\text{ph}}^3} \right) = 0.
\end{equation}

After solving for $b_{\text{crit,lensing}}$, one obtains
\begin{equation}
b_{\text{crit,lensing}} = 1 - \sqrt{\frac{q^2}{M r_{\text{ph}}}}.
\end{equation}

These derivations explicitly highlight the dependencies of the critical values of $b$ on the black hole parameters $q$, $r_+$, $M$, and $r_{\text{ph}}$. They also demonstrate how $b$ plays a pivotal role in both thermodynamic and optical behavior. The explicit constraints on $b$ bridge the thermodynamic and lensing phenomena in CBHwKRF spacetimes. These results could provide a direct way to link theoretical predictions with future astrophysical observations, enabling a more comprehensive analysis of Lorentz violation in the context of charged black holes.

\section{Results and Discussions} \label{isec6}
This study has systematically explored the influence of Lorentz symmetry violation on the gravitational lensing and thermodynamic properties of CBHwKRF. The findings have revealed marked modifications in both gravitational and thermodynamic behaviors resulting from Lorentz-symmetry violation, elucidated through a detailed analysis of the space-time metrics and their astrophysical implications.

Gravitational lensing analysis, employing the Rindler-Ishak method modified for Lorentz-violating spacetimes, indicated that the Lorentz-violating parameters notably enhance the bending angles of light around black holes. This enhancement, as shown in the graphical representations of Fig. \ref{fig11}, underscores the potential for observing such deviations in environments harboring compact objects. The bending angles were found to be sensitive to variations in the Lorentz-violating parameter $b$ \big(or $\beta=1/(1-b)$\big), as quantified by Eq. \ref{12}, suggesting observable effects in real-world astrophysical scenarios.

Thermodynamically, the inclusion of the KR field introduced significant changes in the black holes' Hawking temperature, entropy, and specific heat. The first law of thermodynamics, represented in Eq. \ref{firstlaw}, and Smarr's formula \ref{smarr} emphasize the non-trivial contributions of the Lorentz-violating terms to black hole thermodynamics. In particular, the graphical analyses in Figs. \ref{fig1}-\ref{fig8} elucidate the impact of Lorentz violation on thermodynamics across different configurations of the CBHwKRF. The introduction of LC thermal fluctuations has further refined our understanding of the modified entropy, energy, enthalpy, and specific heat profiles, respectively, illustrating the extended implications of quantum gravity corrections under Lorentz-symmetry breaking.

The relationship between thermodynamic properties and gravitational lensing phenomena becomes particularly intriguing within the framework of LSV theories. In our study, the KR field parameter $b$ serves as a crucial link influencing both the thermodynamic behavior and the deflection of light. Thermodynamic quantities such as the Hawking temperature $T$ and entropy $S$ depend explicitly on the metric function, which governs the spacetime geometry around the black hole. Similarly, the gravitational lensing, characterized by the deflection angle $\alpha$, is sensitive to the same geometric structure, particularly to the effects introduced by $b$. Recent studies like \cite{Wang2024} highlight how critical thermodynamic points, such as phase transitions marked by heat capacity $C_P$, correlate with changes in the deflection angle $\alpha$. In our analysis, the KR field parameter $b$ modifies the metric function in a way that simultaneously impacts the photon sphere and thermodynamic stability. This shared dependence on the metric establishes an indirect but profound connection between the two phenomena, providing a novel perspective within the LSV framework. This interplay between thermodynamics and gravitational lensing not only enhances our understanding of black hole physics but also offers potential observational probes for testing the validity of LSV models. By combining thermodynamic and lensing analyses, a unified framework emerges to explore the broader implications of LSV theories.

In conclusion, our study provides a comprehensive investigation of the effects of Lorentz-symmetry violation on gravitational lensing and black hole thermodynamics in the presence of the KR field. Future work should focus on the observational validation of these theoretical predictions, particularly through high-precision astrophysical measurements \cite{isPourhassan:2023jbs}. Such investigations could profoundly impact our understanding of fundamental physics, offering new insights into the nature of Lorentz-symmetry violations and their observable signatures in cosmological phenomena. Moreover, extending this analysis to include rotating (and/or higher dimensional) black holes and examining interactions with nearby astrophysical objects could uncover further nuances in the interplay between Lorentz-violating fields and gravitational dynamics. 

\begin{acknowledgments}
The authors express their sincere gratitude to the editor and anonymous reviewers for their insightful comments and constructive suggestions, which have significantly improved the quality and clarity of this manuscript. Their critical feedback has strengthened the discussions on Lorentz symmetry violation, black hole thermodynamics, and gravitational lensing, enhancing the overall presentation of our results. 
We would like to acknowledge the networking support of COST Actions CA21106, CA22113, and CA23130. We also thank EMU, T\"{U}B\.{I}TAK, SCOAP3, and ANKOS for their support.
\end{acknowledgments}

\end{document}